\documentclass[aps,prl,twocolumn,preprintnumbers,nofootinbib,showpacs]{revtex4}
\usepackage{graphicx}
\begin{document}
\title{Black hole production and Large Extra Dimensions}
\author{Kingman Cheung}
\email[Email:]{cheung@phys.cts.nthu.edu.tw}
\affiliation{National Center for Theoretical Sciences, National Tsing Hua 
University, Hsinchu, Taiwan, R.O.C.}
\date{\today}

\begin{abstract}
Black hole (BH) production at colliders is possible when the colliding energy 
is above the Planck scale, which can effectively be at TeV scale in models of
large extra dimensions.  In this work, we study the production of black holes
at colliders and discuss the possible signatures.  We point out the 
``$ij\to \;{\rm BH}+ $ others'' subprocesses, in which the BH and other
SM particles are produced with a large transverse momentum.  When the BH
decays, it gives a signature that consists of particles of high multiplicity
in a boosted spherical shape on one side of the event and a few number of
high $p_T$ partons on the other side, which provide very useful tags for 
the event.
\end{abstract}
\pacs{}
\preprint{NSC-NCTS-011012}
\maketitle

{\it Introduction.}
Black hole (BH) has been an illusive object for decades, as we cannot directly
measure any properties of it, not to mention the production of black holes in
any terrestrial experiments.  This is due to the fact that in order to
produce black holes in collider experiments one needs a center-of-mass energy
above the Planck scale ($M_P\sim 10^{19}$ GeV), which is obviously inaccessible
at the moment.  

Since the second revolution of string theories, a crop of models with
extra dimensions have been proposed to solve various theoretical problems.
In an attractive 
 model of large extra dimensions or TeV quantum gravity \cite{arkani},
the effective Planck scale can be as low as a few TeV.  
This is made possible by localizing the standard model (SM) particles on a
brane (using the idea of D-branes in Type I or II string theory),
while gravity is free to propagate in all dimensions.

The fact that the effective Planck scale is as low as TeV opens up an
interesting possibility of producing a large number of black holes
\cite{hole,banks,emp} at collider experiments (e.g. LHC).
Reference \cite{emp} also showed that a BH localized on a brane will 
radiate mainly in the brane, instead of radiating into the 
Kaluza-Klein states of gravitons of the bulk. 
In this case, the BH so produced will decay mainly into the
SM particles, which can then be detected in the detector.  This opportunity
has enabled investigation of the properties of BH at terrestrial collider
experiments.

A black hole is characterized by its Schwarzschild radius $R_{\rm BH}$, which
depends on the mass $M_{\rm BH}$ of the BH.  A simplified picture for 
BH production is as 
follows.  When the colliding partons have a center-of-mass (c.o.m.) energy
above some thresholds of order of the 
Planck mass and the {\it impact parameter} less 
than the Schwarzschild radius $R_{\rm BH}$, a BH is formed
and almost stationary in the c.o.m. frame.  The BH so produced will decay
thermally (regardless of the incoming particles) and thus isotropically 
in that frame. 

This possibility has been investigated in a few recent work at the LHC
\cite{scott,greg,hoss,volo}, and at cosmic ray experiments 
\cite{feng,anch,ratt}. 
In Refs. \cite{scott,greg,hoss} black hole production in hadronic collisions
is calculated in $2\to 1$ subprocesses: $ij\to\; {\rm BH}$, where $i,j$ are
incoming partons.  The black hole so produced is either at rest or traveling along 
the beam pipe such that its decay products (of high multiplicity)
have a zero net transverse momentum ($p_T$).  
Dimopoulos and Landsberg \cite{greg} demonstrated that a BH so produced 
will decay with a high multiplicity \cite{greg}.
Banks and Fischler \cite{banks} pointed out that one of the most striking
signature of BH production would be a complete cutoff of process with 
$p_T$ larger than $R_{\rm BH}^{-1}$ because the incoming partons can never 
get close enough to perform an extremely hard QCD scattering.  

In this Letter, we point out the ``$ij\to {\rm BH}+\;{\rm others}$''
 subprocesses,
such that the BH is produced with a large $p_T$ before it decays.
In such subprocesses, the ``others'' are just the ordinary SM particles and
usually of much lower multiplicity than the decay products of the BH.
Therefore, the signature would be very striking: on one side of the event
there are particles of high multiplicity (from the decay of the BH), the
total $p_T$ of which is balanced by a much fewer number of particles on 
the other side.  Such a signature is very clean and 
should have very few backgrounds.  

The ``$ij\to {\rm BH}+\;{\rm others}$'' subprocesses
can be easily formed when the c.o.m. energy of the colliding particles 
are larger than the BH mass, the 
excess energy will be radiated as other SM particles.  Thus, a picture of
$ij\to {\rm BH}+\;{\rm others}$ is justified.
Besides, such processes are of immense interests 
because they involve qualitatively new phenomena, e.g., violation of
global quantum (lepton, baryon) numbers.
The resulting cross section is large enough for detection.
Even at $p_T>500$ GeV the production 
cross section is as large as 24 fb for $n=4,M_D=1.5$ TeV, and 
$y\equiv (M_{\rm BH})_{\rm min}/M_D=5$ (a BH with a mass $\ge 7.5$ TeV),  
which gives $2400$ clean events
for an integrated luminosity of 100 fb$^{-1}$ at the LHC, as will be shown
later.  The event rate is still large enough for detection.  Here we have 
imposed a very restricted requirement on the BH entropy 
($S_{\rm BH} \agt 25$) to ensure the validity of the classical 
BH description.  Otherwise, if we relaxed this entropy constraint 
the production cross sections would be increased tremendously, but the
cross sections have to be interpreted with care because of the 
presence of large string effects in this regime.

{\it Production.}
The Schwarzschild radius $R_{\rm BH}$ of a BH of mass $M_{\rm BH}$ in
$4+n$ dimensions is given by \cite{myers}
\begin{equation}
\label{r}
R_{\rm BH} = \frac{1}{M_D}\; \left (
\frac{M_{\rm BH}}{M_D} \right)^{\frac{1}{n+1}}\; 
\left( \frac{ 2^n \pi^{ \frac{n-3}{2}} \Gamma(\frac{n+3}{2} )}{n+2} 
\right )^{\frac{1}{n+1}} \;,
\end{equation}
where $M_D$ is the effective Planck scale in the model of large extra
dimensions defined by
\begin{equation}
M_D^{n+2} = \frac{ (2\pi)^n}{8 \pi G_{4+n} } \;,
\end{equation}
where $G_{4+n}$ is the gravitational constant in $D=4+n$ dimensions 
(used in the Einstein equation: ${\cal R}_{AB} - \frac{1}{2} g_{AB}{\cal R} = 
- 8 \pi G_{4+n} T_{AB}$.)
The radius is much smaller than the size of the extra dimensions.
BH production is expected when the colliding partons with a center-of-mass
energy $\sqrt{\hat s} \agt M_{\rm BH}$ pass within a distance less than 
$R_{\rm BH}$.  A black hole of mass $M_{\rm BH}$ is formed and the rest of 
energy, if there is, is radiated as ordinary SM particles.   
This semi-classical argument calls for a geometric approximation for 
the cross section for producing a BH of mass $M_{\rm BH}$ as
\begin{equation}
\label{geo}
\sigma(M_{\rm BH}^2 ) \approx \pi R_{\rm BH}^2 \;.
\end{equation}
In the $2\to 1$ subprocess, the c.o.m. energy of the colliding partons is just
the same as the mass of the BH, i.e., $\sqrt{\hat s}=M_{\rm BH}$, which
implies a subprocess cross section
\begin{equation}
\hat \sigma(\hat s) = \int \; d\left(\frac{M_{\rm BH}^2}{\hat s}\right )\; 
\pi R_{\rm BH}^2 \; \delta\left( 1 - M^2_{\rm BH}/\hat s \right ) 
= \pi R_{\rm BH}^2 \;.
\end{equation}
On the other hand, for the $2\to n (n\ge 2)$ subprocesses the subprocess
cross section is 
\begin{equation}
\hat \sigma(\hat s) = \int^{1}_{ (M^2_{\rm BH})_{\rm min}/\hat s} \; 
d\left(\frac{M_{\rm BH}^2}{\hat s}\right )\; \pi R_{\rm BH}^2 \;.
\end{equation}

Another important quantity that characterizes a BH is its entropy given by
\begin{equation}
\label{entropy}
S_{\rm BH} = \frac{4\pi}{n+2}\; \left (
\frac{M_{\rm BH}}{M_D} \right)^{\frac{n+2}{n+1}}\; 
\left( \frac{ 2^n \pi^{ \frac{n-3}{2}} \Gamma(\frac{n+3}{2} )}{n+2} 
\right )^{\frac{1}{n+1}} \;.
\end{equation}
To ensure the validity of the above classical description of 
BH \cite{giddings},
the entropy must be sufficiently large, of order 25 or so.
We verified that that when $M_{\rm BH}/M_D \agt 5$, the entropy 
$S_{\rm BH}\agt 25$, below that string effects are important.
Therefore, to avoid getting into the nonperturbative regime of the BH and 
to ensure the validity of the semi-classical formula, we restrict the mass 
of the BH to be $M_{\rm BH}> y M_D$, where 
$y \equiv (M_{\rm BH})_{\rm min}/M_D$ is of order 5.

Voloshin \cite{volo} pointed out that the semi-classical argument for the BH
production cross section is not given by the geometrical cross section
area, but, instead,  suppressed by an exponential factor:
\begin{equation}
\label{supp}
%\exp \left( - \left[ \frac{M_{\rm BH}}{M_D} \right ]^{\frac{n+2}{n+1}} 
\exp \left( - \frac{S_{\rm BH}}{n+1} \right ) \;.
\end{equation}
The suppression factor makes the production of BH concentrate on
$M_{\rm BH}$ close to $M_D$.  Thus, if the available energy in the
collision is larger than the $M_D$ the rest of energy is more likely to 
radiate as the SM particles. 
There are, however, counter arguments \cite{giddings,emparan} that
the simple geometric formula should be valid. 
\footnote{
There is also a counter-counter argument from Voloshin \cite{volo2}.
Nevertheless, before the issue is resolved we present the results with
and without the suppression factor.}

\begin{figure}[th!]
\includegraphics[width=3in]{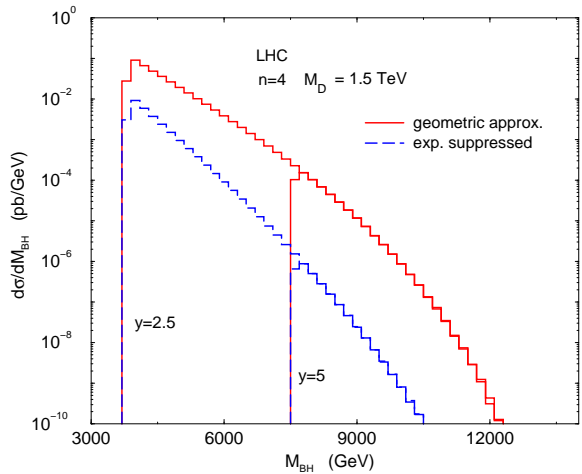}
\caption{\small
\label{mbh}
The differential cross section $d\sigma/d M_{\rm BH}$ for $pp \to 
{\rm BH} + 1$ parton versus the mass $M_{\rm BH}$ at the LHC
for $n=4, M_D=1.5$ TeV and $y\equiv (M_{\rm BH})_{\rm min}/M_D=2.5, 5$.
``Geometric approx.'' means that we used Eq. (\ref{geo}) to
calculate the cross section while ``exp. suppressed'' means that we also 
included the suppression factor in Eq. (\ref{supp}).
}
\end{figure}

In this work, we first consider both forms of cross sections: (i) the naive
$\pi R_{\rm BH}^2$ and (ii) the $\pi R_{\rm BH}^2$ multiplied
with the exponential factor of Eq. (\ref{supp}).  
But we shall see immediately that the suppression factor renders 
the cross section to be too small for detection.
In Fig. \ref{mbh}, we show the differential cross section
$d\sigma/dM_{\rm BH}$ for the process $pp\to {\rm BH}+1$ parton at the 
LHC for $n=4$, $y \equiv (M_{\rm BH})_{\rm min}/M_D =5$ and $M_D=1.5$ TeV,
which is consistent with the existing limit on $M_D$ \cite{limits}.  
We can see that when the exponential suppression factor is used
the spectrum of $M_{\rm BH}$ will shift closer to $M_D$.  
The average value $\langle M_{\rm {BH}} \rangle$ for the
geometric approximation is about $8.0$ TeV while using the exponential
suppression factor the $\langle M_{\rm {BH}} \rangle \simeq 7.8$ TeV. 
However, the cross section is suppressed 
more than two orders of magnitude relative to the naive geometric cross 
section. From here on we shall not concern this suppression factor anymore.
We also show the graphs for $y=2.5$, a less stringent 
requirement on the BH entropy.  Obviously, the production cross section
is much larger.  However, careful interpretation is needed because in this
BH mass region it might involve nonperturbative string corrections.

\begin{figure}[bh!]
\includegraphics[width=3in]{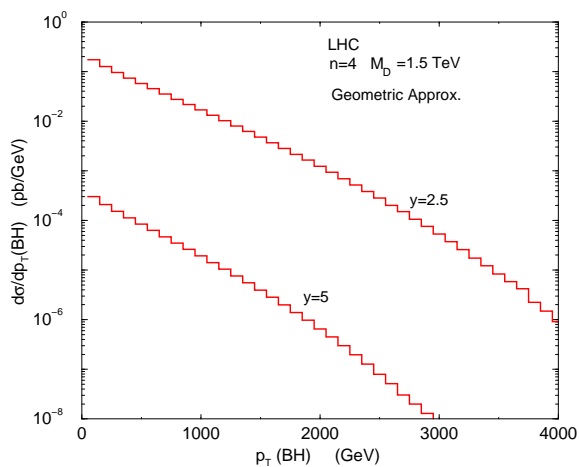}
\caption{\small
\label{pt}
The transverse momentum spectrum for the BH production in association with
a massless parton at the LHC for $n=4, M_D=1.5$ TeV and
$y\equiv (M_{\rm BH})_{\rm min}/M_D=2.5,\, 5$.}
\end{figure}

The main difference between the $2\to 1$ and $2\to n\;(n\ge 2)$ subprocesses
is that in $2\to n$ the 
BH will have a transverse momentum.  We approximate the $2\to n$ 
subprocess by a $2\to 2$ subprocess and assume the BH is produced in 
association with a massless parton.  Since the mass of the BH is larger than
the effective Planck mass, the c.o.m. energy $\sqrt{\hat s} \ge M_{\rm BH}$ 
is of order of a few 
TeV and would give a large transverse momentum to the BH.
We show in Fig. \ref{pt} the transverse momentum spectrum for the
production $pp \to {\rm BH}\; + 1$ parton.  
The average value of $p_T$ is about $330$ GeV for $n=4,M_D=1.5$ TeV and 
$y=5$.

{\it Decay.}
The main phase of the decay of BH is via the Hawking evaporation.  The 
evaporation rate is governed by its Hawking temperature, which is
given by \cite{myers}
\begin{equation}
T_{\rm BH} = \frac{n+1}{4\pi R_{\rm BH} }\;
\end{equation}
which scales inversely with some powers of $M_{\rm BH}$.  The heavier the
BH the lower the temperature is.  Thus, the evaporation rate is slower.
The lifetime of the BH scales inversely with the Hawking temperature 
as given by
\begin{equation}
\tau \sim \frac{1}{M_D} 
\left( \frac{M_{\rm BH}}{M_D} \right )^{\frac{n+3}{n+1}} \;.
\end{equation}
{}From the above equation, it is obvious to see that the lifetime of a BH
becomes much longer in models of large extra dimensions than in the usual
$4D$ theory.  However, the lifetime is still so short that it will decay
once being produced and no displaced vertex can be seen in the detector.
For another view point on the BH decay please see Ref.~\cite{casa}.

\begin{figure}[th!]
\includegraphics[width=3in,angle=-90]{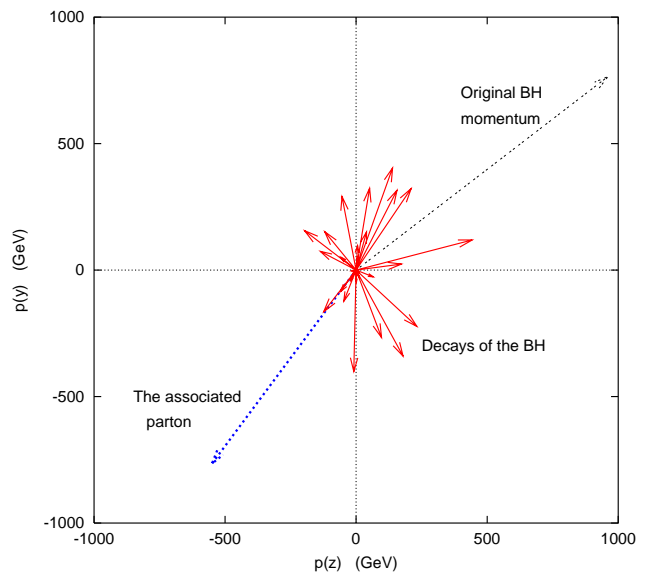}
\caption{\small
\label{cartoon}
A cartoon showing a typical event of black hole production with a large
transverse momentum.}
\end{figure}

Another important property of the BH decay is the large number of particles,
in accord with the large entropy in Eq. (\ref{entropy}), 
in the process of evaporation.  
Dimopoulos and Landsberg \cite{greg}
showed that the average multiplicity $\langle N \rangle$ in the decay of a BH
is order of $10-30$ for $M_{\rm BH}$ being a few times of
$M_D$ for $n=2-6$.  
Since we are considering the BH that has an entropy of order 25 or more, it 
guarantees a high multiplicity BH decay.
The BH decays more or less isotropically and
each decay particle has an average energy of a few hundred GeV.  Therefore,
if the BH is stationary, the event is very much like a spherical event with
many particles of hundreds of GeV pointing back to the interaction point.
Moreover, the ratio of hadronic to leptonic activities in the BH decay is
about $5:1$ \cite{scott}.
On the other hand, if the BH is produced in association with other SM
particles (as in $2\to n$ subprocesses), the BH decay will be a boosted
spherical event on one side, 
the transverse momentum of  which is balanced 
by a few number of particles on the other side.
A cartoon for such a typical event in the $(y,z)$ plane is shown in 
Fig. \ref{cartoon}.  This is a high $p_T$ event.  On one side of the event
is the decay products of high multiplicity of the BH in a boosted spherical
shape.  The original momentum of the BH in the $(y,z)$ plane 
is also shown, which
is balanced by the momentum of the energetic parton, which is on the other 
side of the event.  Such spectacular events should have negligible background.

In addition, since at the LHC multi-parton collisions and 
overlapping events may be likely to happen, a careful discrimination is 
therefore necessary, especially,
in the case that the BH is produced at rest or moving along the beam-pipe
(i.e. in $2\to 1$ subprocess).  In our study, the $2\to 2$ subprocess affords
an easier signature experimentally.  The high $p_T$ parton emerging as a jet,
a lepton, jets, or leptons provides an easy tag.

Another concern of BH production is the event rate because the higher the $p_T$
the smaller the cross section is.  
The production cross section for $p_T>500$ GeV is as large as 24 fb for 
$n=4, M_D=1.5$ TeV and $y\equiv (M_{\rm BH})_{\rm min}/M_D=5$,  
which corresponds to $2400$ events for 
an integrated luminosity of 100 fb$^{-1}$. 
Such a large number of clean events should be observable at the LHC.  
The production cross sections for other values of $n$, $M_D$ and $y$
are listed in Table \ref{table1}.
The cross sections listed for $y\equiv (M_{\rm BH})_{\rm min}/M_D \alt 4$ 
should be interpreted with care, because the smaller the ratio
$(M_{\rm BH})_{\rm min}/M_D$ the stronger the string effect is and the
classical description for BH may not be valid.  Nevertheless, the numbers
listed here can be used for comparison with other published results.

In this paper, we have emphasized the importance and the advantages of using
the $2\to 2$ subprocess for BH production, which allows a substantial 
transverse momentum kick to the BH, and at the same time produce an 
energetic high $p_T$ parton, which provides a critical tag to the event.
The observation here serves as an interesting extension to the previous
work, in which the consideration is only given to BH production with
the BH at rest. 

This research was supported in part by the National Center for Theoretical
Science under a grant from the National Science Council of Taiwan R.O.C.

\begin{table*}[bh!]
\caption{\small \label{table1}
Cross sections in pb for BH production in ``$2 \to 2$'' 
subprocess for various values of $n$, $M_D$ and 
$y \equiv (M_{\rm BH})_{\rm min}/M_D$ at the LHC.  
The transverse momentum cut is $p_T>500$ GeV.  The cross 
sections are calculated using the geometric approximation of Eq. (\ref{geo}).
The cross sections given in the parenthesis are for the ``$2\to 1$'' 
subprocess, i.e., with zero $p_T$ and the BH at rest.
For $M_D=3$ TeV and $y=5$ the minimum mass for the BH is already larger
than the energy of the LHC.
}
\medskip
\begin{ruledtabular}
\begin{tabular}{cllll}
                          & \multicolumn{4}{c}{Cross section in pb} \\
                          & $n=3$ &  $n=4$ &  $n=5$&   $n=6$ \\
\hline
\underline{$M_D=1.5$ TeV} &       &        &       &  \\
  $y=1$                   & 351 (5300)   & 571 (8650)   & 820 (12400)  & 
1090 (16600) \\
  $y=2$                   & 40.1 (540) & 62.8  (831) & 87.1 (1150) & 113 
      (1490)           \\
  $y=3$                   & 4.2  (70.4) &  6.3 (105)  & 8.6  (142) & 10.9 (180) \\
  $y=4$                   & 0.34 (8.1) &  0.49 (11.8) & 0.65 (15.7) & 0.82 (19.8) \\
  $y=5$                   & 0.017 (0.68) &  0.024 (0.97)& 0.032 (1.3)& 0.039 (1.6)  \\
\hline
\underline{$M_D=2$ TeV} &  &  &   &  \\
  $y=1$                   &  83.9 (1120)  & 137 (1840) & 198 (2650) & 264 (3540)  \\
  $y=2$                   &  4.5  (67.8)  & 6.9 (105) & 9.7 (145)& 12.5 (188)\\
  $y=3$                   &  0.16 (4.0)  & 0.25 (5.9)&0.33 (8.0)& 0.43 (10.3)\\
  $y=4$                   &  0.0026 (0.13) & 0.0038 (0.19)& 0.0051 (0.25) & 
                             0.0064 (0.32) \\
  $y=5$                   &  $7\times 10^{-6}$ (0.0012) & 
                             $1.1\times 10^{-5}$ (0.0017) &   
                             $1.4\times 10^{-5}$ (0.0022) & 
                             $1.7\times 10^{-5}$ (0.0028)  \\
\hline
\underline{$M_D=3$ TeV}   &      &  &   &  \\
  $y=1$                   & 7.2 (95.5) & 11.9 (157) & 17.3 (228)  &23.1 (305)\\
  $y=2$                   & 0.06 (1.4)& 0.09 (2.2) & 0.13 (3.1) &  0.17 (4.1)\\
  $y=3$                   & $0.69\times 10^{-4}$ (0.0056)& 
         $1.0\times10^{-4}$ (0.0084)& $1.4\times 10^{-4}$ (0.011) & 
           $1.8\times 10^{-4}$ (0.015)\\
  $y=4$                   & $7\times 10^{-11}$ ($1.8\times 10^{-7}$) & 
      $1.1\times 10^{-10}$ ($2.6\times 10^{-7}$) & 
                            $1.4\times 10^{-10}$ ($3.5\times 10^{-7}$) & 
                  $1.8\times 10^{-10}$ ($4.5\times 10^{-7}$)  \\
  $y=5$                   & -  &  -& -  &  - \\
\end{tabular}
\end{ruledtabular}
\end{table*}

%%%%%%%%%%%%%%%%%%%%%%%%%%%%%%%%%%%%%%%%%%%%%%%%%%%%%%%%%%%%

%%%%%%%%%%%%%%%%%%%%%%%%%%%%%%%%%%%%%%%%%
\end{document}